\documentclass[preprint,numbers,nocopyrightspace]{sigplanconf}

\usepackage[latin9]{inputenc}
\usepackage{color}
\usepackage{array}
\usepackage{multirow}
\usepackage{amsmath}
\usepackage{amssymb}
\usepackage{graphicx}
\usepackage{epstopdf}
\usepackage{algorithm}
\usepackage{flushend}
\usepackage{url}
\usepackage{hyperref}

\usepackage[letterspace=-100]{microtype}

\usepackage{color}
\usepackage[table]{xcolor}
\definecolor{azure}{RGB}{51, 102, 153}
\definecolor{light}{RGB}{224, 224, 224}
\definecolor{lightblue}{RGB}{204, 204, 255}
\usepackage[font=small]{caption}

\makeatother

\begin{document}

\setlength{\pdfpageheight}{\paperheight}
\setlength{\pdfpagewidth}{\paperwidth}

%\conferenceinfo{HLPC '16}{April 2, 2016, Atalanta, USA}
%\copyrightyear{2016}
%\copyrightdata{978-1-nnnn-nnnn-n/yy/mm}
%\copyrightdoi{nnnnnnn.nnnnnnn}

%\titlebanner{banner above paper title}    % These are ignored unless
%\preprintfooter{short description of paper}  % 'preprint' option specified.	

\title{On the Synchronization of Intermittently Powered Wireless Embedded Systems}

%\authorinfo{Blinded for Review}{}{}

\authorinfo{Kas{\i}m Sinan Y{\i}ld{\i}r{\i}m \and Henko Aantjes \and Amjad Yousef Majid \and Przemys{\l}aw Pawe{\l}czak}
{Embedded Software Group, Delft University of Technology, Mekelweg 4, 2628 CD Delft, The Netherlands}
{\{k.s.yildirim, a.y.majid, p.pawelczak\}@tudelft.nl, h.aantjes@student.tudelft.nl}

\maketitle

\begin{abstract}
	
Battery-free computational RFID platforms, such as WISP (Wireless Identification and Sensing Platform), are emerging intermittently powered devices designed for replacing existing battery-powered sensor networks. As their applications become increasingly complex, we anticipate that synchronization (among others) to appear as one of crucial building blocks for collaborative and coordinated actions. With this paper we aim at providing initial observations regarding the synchronization of intermittently powered systems. In particular, we design and implement the first and very initial synchronization protocol for the WISP platform that provides explicit synchronization among individual WISPs that reside inside the communication range of a common RFID reader. Evaluations in our testbed showed that with our mechanism a synchronization error of approximately 1.5 milliseconds can be ensured between the RFID reader and a WISP tag.

\end{abstract}



\section{Introduction}\label{sec:Intro}

Low-power wireless embedded systems consist of tiny, low-cost, battery-operated and spatially separated computers that communicate with eachother through wireless medium by exchanging radio packets. Powering these small-scale embedded systems, e.g. wireless sensor networks (WSNs), is still a crucial problem~\cite{emergence}. Replenishment or recharging their batteries are impractical and inhibit long-term operation. Moreover, batteries increase the size and cost of their hardware. Fortunately, the energy efficiency of these systems has improved considerably such that their power requirements are in the order of a few $\mu$W \cite{booksmith2013}. Furthermore, recent advancements in microelectronics technology enabled harvesting power from radio frequency (RF) sources which is sufficient to power low-power embedded systems in practice~\cite{60ghz}. Nowadays, the growth of RF-powered computing paradigm is bringing new research opportunities and challenges~\cite{emergence}, leading to a promising class of low-power embedded systems, so called Intermittently Powered Devices (IPDs).

By taking existing RFID (Radio Frequency IDentification) technology as a foundation, computational RFIDs (CRFIDs) are emerging IPDs that allow sensing, computation and communication without batteries---replacing existing battery-powered sensor networks~\cite{booksmith2013}. CRFIDs are equipped with a backscatter radio composed of a simple circuitry that modulates the carrier wave generated by a reader to transmit information. This allows communication to come almost for free, which is a fundamental difference from sensor networks as the transceiver circuit is the most energy-hungry component in WSN. In CRFID domain, the bottleneck in terms of power consumption has shifted from communication to computation and sensing~\cite{ekhonet}. A typical example of CRFID systems is the WISP (Wireless Identification and Sensing Platform)~\cite{wisp}. Commercial RFID readers implementing EPC Gen2
standard~\cite{epc_gen2} are used to provide power to WISP tags and to receive data from them. Apart from low data rate sensing~\cite{neuralWISP}, these maintenance-free devices are evolving to support also high data rate, more complex and richer sensing applications such as continuous sensor data-streaming, e.g. capturing and transfer of images using battery-free cameras, i.e. WISPCam~\cite{wispcam_2015,cameralocalize_2015}.

\subsection{Motivation and Challenges}

As CRFID platforms and applications are developing, these systems are 
anticipated to demand their own implementation of basic sensor network building 
blocks. As an example, consider a hypothetical application of multiple WISPCams 
that are deployed to capture images of an object from different angles 
simultaneously. Since each WISPCam is spatially distributed and has its own 
clock hardware whose oscillator generate pulses at slightly different speeds, 
they require a \emph{synchronization} service to obtain a common time notion 
for such collaborative and coordinated actions. However, the characteristics of 
CRFID systems expose fundamentally different challenges than sensor networks to 
implement these services. The reason is mainly twofold: 

\begin{itemize}
\item \textbf{Challenge I:} CRFID systems should perform computations in an 
energy efficient manner despite of intermittent RF power that leads to loss of 
computational state frequently, e.g. when RFID reader moves away from the CRFID 
tag \cite{ransford:2008}.

\item \textbf{Challenge II:} Continuously varying voltage supply introduces 
severe hardware instability, e.g. varying oscillator frequencies in short-term 
that effects the stability of the clocks, leading to degraded accuracy of 
computation and sensing.	
\end{itemize}
 
\subsection{Contributions}

Considering these facts, our focus is to answer the question of \emph{How to 
design a building block that synchronizes intermittently powered wireless 
embedded systems?} To this end, first we investigate the WISP platform and 
provide initial observations and limitations pertaining to the synchronization 
of these devices. Even though there are studies focused on the synchronization 
of multiple RFID readers in the current literature, e.g. \cite{reader_sync}, we 
are unaware of any existing study that provides explicit synchronization among 
individual WISPs that reside inside the communication range of a common RFID 
reader. Hence, our main contribution is to design and implement the first and 
very initial \emph{reader-tag} synchronization primitive for the WISP platform. 
Evaluations in our testbed showed that a maximum synchronization error of 
approximately 1.5 milliseconds can be ensured between the reader and a WISP tag 
with this mechanism.

\section{Synchronization in Conventional Wireless Sensor Networks}
\label{sec:WSNs}

The instability of the clock hardware, delays during communication among sensor nodes and software methods to establish synchronization are the main points effecting the synchronization in conventional WSNs. 

\textbf{Clock Hardware:} In WSNs, each sensor node is equipped with a built-in clock that is implemented as a counter register clocked by a low-cost external crystal oscillator. At each oscillator pulse, i.e. \emph{tick} of the clock, the counter register is incremented. The duration between two consecutive ticks is the \emph{rate} of the built-in clock. Environmental factors such as temperature, voltage level and aging of the crystal prevent built-in clocks to generate ticks at the exact speed of real-time, leading to bounded \emph{clock drift}. The prominent environmental factor affecting the frequency of the built-in clocks is the temperature~\cite{Lenzen2009Optimal}. Moreover, \emph{quantization errors} occur with low-frequency built-in clocks, that prevents precise timing measurements. 

\textbf{Wireless Communication:} Due to their different clock frequencies, sensor nodes exchange their clock information periodically to synchronize their built-in clocks by computing a \emph{software clock} that represents synchronized notion of time. The difference between the reference time and the software clock is the \emph{synchronization error}. In WSNs, the synchronization error is mainly affected by several sources of errors. Delays introduced during the wireless communication between participating nodes is the major error source. The \emph{transmission delay}, defined as the time that passes between the start of the broadcast attempt and the receipt by the receiver node, is composed of deterministic and non-deterministic components~\cite{Maroti2004}. Assigning timestamps at the MAC layer is a common method in WSNs to get rid of the deterministic delay components and to improve synchronization accuracy. This obligates radios like Chipcon CC2420 which allows assignment of time information to a radio packet just before transmission and reception. 

\textbf{Computation Methods:} Due to the multi-hop nature of WSNs, the most common synchronization mechanism is to propagate the time information of a particular reference node to let receiver nodes synchronize themselves to the received reference time information by employing \emph{least-squares regression}~\cite{Maroti2004,Lenzen2009Optimal}. There are also fully-distributed approaches in which sensor nodes interact only with their direct neighbours in a peer-to-peer fashion and employ methods based on \emph{distributed consensus} \cite{Sommer2009Gradient,SchenatoFiorentin:2011}. Since the transceiver is the most power-hungry circuit, the objective of synchronization in WSNs is generally to reduce re-synchronization frequency to decrease communication overhead and save power.

\section{\label{sec:CRFIDs}Fundamental Challenges of Synchronizing IPDs: The CRFID Case}

After presenting a brief summary of the synchronization in WSNs, we now delve into to the synchronization characteristics of IPDs. To this end, we will consider WISP \cite{wisp}, the de facto CRFID platform. The main aspects pertaining to synchronization in WISP can be stated as follows:

\begin{itemize}
\item \textbf{Single-hop reader-tag architecture:} WISPs are deployed inside the communication range of an RFID reader and they can communicate only with the reader using backscatter communication. Therefore, the RFID reader itself is the natural reference device to establish synchronization among the WISPs, promoting \emph{reader-tag synchronization}. Since WISPs are unable to communicate with their neighboring nodes directly, \emph{tag-tag synchronization} is more challenging.
	
\item \textbf{Continuously varying voltage level:} In WSNs, the battery level decreases gradually that allows stable voltage levels in short-term. On the contrary, fluctuating input voltage prevents short-term stability of the clock hardware and introduces significant drift. Hence, the prominent factor affecting the frequency of the crystal oscillator is the varying voltage level rather than the temperature for WISPs.
	
\item \textbf{Frequent loss of synchronization state:} On the contrary to sensor nodes, WISP tags frequently ``die'' due to power loss and they need to save synchronization state, e.g. data regarding to software clock, into the non-volatile memory to recover when they find sufficient energy. However, saving computational state to non-volatile memory is also an energy consuming task~\cite{blisp}.
	
\item \textbf{Computation and memory overhead sensitivity:} Classical motto of WSNs, ``computation instead of communication whenever possible''~\cite[p. 44]{karl2007protocols}, is no longer valid for WISP platform since backscatter communication comes almost for free~\cite{ekhonet}. We require lightweight methods in terms of computation and memory for the synchronization due to the intermittent power limitations. Since methods like least-squares regression is computationally heavy and require considerable amount of memory~\cite{pi2015}, so they should be avoided. 
	
\item \textbf{Limitations of EPC Gen 2 standard:} WISP firmware implements the EPC Gen 2 standard~\cite{epc_gen2} that increases the compatibility with the existing RFID systems. However, the standard introduces limitations, e.g. currently it does not assign timestamps to the radio packets, which is a fundamental requirement to establish synchronization. Moreover, communication delays between the reader and tag are quite dependent on the implementation of this standard by RFID readers. Unfortunately, these issues lead to less accurate synchronization as compared to existing WSN solutions, as justified by our measurements presented in the following sections. 
\end{itemize}

\section{Reader-Tag Synchronization for WISP Tags}
\label{sec:Reader-Tag}

In this section, we provide initial observations, design and implementation of two reader-tag synchronization approaches and their evaluation in our testbed. We first provide a sender-receiver based synchronization design and present its limitations. Then, we introduce an event-based synchronization mechanism and provide its advantages over the former approach. 

\subsection{Hardware Related Issues and Experimental Testbed}

Before delving into these issues, we first present a brief information about the clock hardware of WISP platform, our testbed setup and implementation details. 

\subsubsection{WISP Clock Hardware}

WISP 5.0 platform comes with MSP430FR5969 \cite{msp430fr5969_data_sheet} microcontroller with FRAM non-volatile memory for ultra low energy data storage and retrieval. The MSP430 clock system supports a 32\,kHz external crystal oscillator, an internal very-low-power low-frequency oscillator, an integrated internal digitally controlled oscillator (DCO). The clock system includes auxiliary clock (ACLK) signal that can be sourced from the external 32\,kHz oscillator. The MSP430 microcontroller has five built-in 16-bit timers that can be clocked with ACLK. Moreover, MSP430 has one active mode and seven software selectable low-power operation modes. In low-power operation mode LPM3, ACKL is active. 

\subsubsection{Testbed Setup}

Our testbed is composed of an RFID reader, a host computer to control this reader and a single WISP tag placed at a university office. We used 915\,MHz Impinj Speedway R1000 RFID reader with firmware version 3.2.4 connected to a Laird S9028PCR 8.5\,dBic gain antenna. We placed the WISP tag at approximately 10 cm from the antenna with line-of-sight and performed all of our experiments with people inside the office. For host-reader control operations, we used sllurp~\cite{sllrp_github}, a LLRP (Low-Level Reader Protocol)~\cite{llrp} control library written in Python. To program the WISP tag, we used a MSP430 Flash Emulation Tool (FET), in combination with TI Code Composer Studio (CCS), attached to the host. In order to explore and characterize the communication between the RFID reader and tag, we used USRP 210 software defined radio, another Laird antenna placed at 50 cm from the tag and GNU Radio~\cite{gnuradio} toolkit to sniff the radio packets during backscatter communication. This allowed us to measure the communication delays on the order of microseconds.

\subsubsection{Software Implementation}

We configured Timer B0 so that its clocked with ACKL and we allowed WISP to transition to LPM3 mode when its idle, thus allowing us to have an 16-bit running timer in low-power mode operation at 32\,kHz precision. WISP-side implementations are done using C and MSP430 assembly.\footnote{We note that all scripts used to generate the results in the paper (including parsing, post-processing and measurement results) are available upon request% or via  \href{https://goo.gl/YlSF7Q}{https://goo.gl/YlSF7Q}.}
.}
\subsection{Notation}\label{subsec:notation}

The timer running at 32\,kHz can be considered as the built-in clock, i.e. \emph{local clock}, of the WISP. This local clock is denoted by $C_{w}$ and its value at any real time $t$ can be modeled as $C_{w}(t) = \intop_{t_0}^{t}f_{w}(\tau)d\tau$. Here, $t_0$ represents the time at which WISP powered on and  $f_{w}(\tau)$ represents the \emph{frequency} of the local clock at time $\tau$. Since the crystal oscillators have bounded drifts, e.g. typically a nominal value $f_{nom}$ is known together with a lower bound $f_{\min}$ and an upper bound $f_{\max}$, we assume that $f_{\min}\leq f_{w}(\tau)\leq f_{\max}$ holds. We denote the clock of the RFID reader by $C_{r}(t)$, which is considered as the reference time for the WISP tag. By collecting reference time information, the WISP tag can build a \emph{software clock}, denoted by $S_{w}$, whose value at any real time $t$ represents the synchronized notion of time. The objective of synchronization is to minimize the \emph{synchronization error} at any time $t$, which is formally defined as $\gamma(t)=S_w(t)-C_r(t)$. 

\subsection{Approach 1: Sender-Receiver Based Synchronization} \label{subsec:sender_receiver}

\begin{figure}
	\centering
	\includegraphics[scale=0.8]{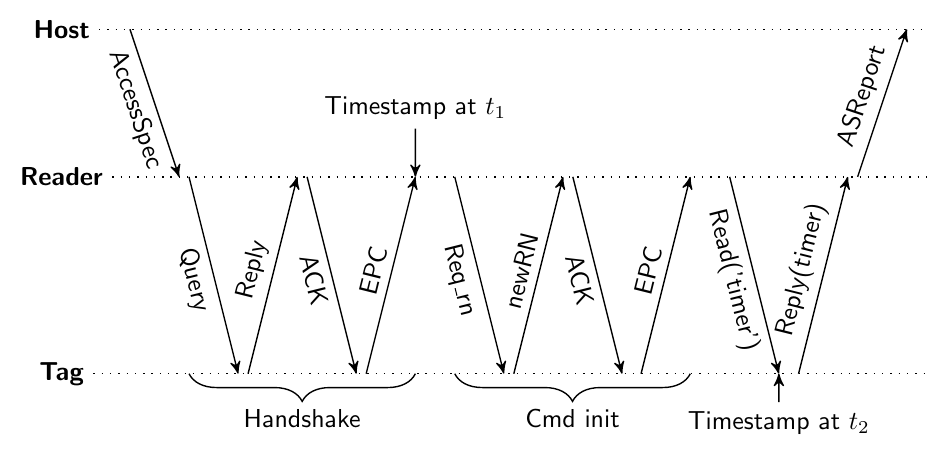}
	\caption{\label{fig:sender_receiver} The message exchange among the host computer, the RFID reader and the tag for sender-receiver synchronization. The host machine sends the high level commands to the reader via \emph{AccessSpec} message and receives the results through an \emph{ASReport}. The reader follows the steps defined in EPC Gen2 standard: (i) performs the \emph{Handshake} to initialize the communication with the active tag; (ii) performs the \emph{Command Initialization} by requesting a random number from the tag (Req\_n) and receiving the random number (newRN); (iii)  performs a \emph{Read} command by sending its request and receiving the timer value. The reader assigns the \emph{FirstSeenTimestamp} to the tag at time $t_1$ and the tag timestamps the command reception event at time $t_2$ with its local clock reading.}
\end{figure}

In sender-receiver based synchronization mechanisms, receiver devices synchronize to the clock of a reference sender device. In order to synchronize itself to the RFID reader with such a mechanism, the WISP tag should obtain several $(C_w(t),C_r(t))$ synchronization points and establish a relationship between its local clock $C_w$ and the reader clock $C_r $, represented by its software clock $S_w$. The value $S_w(t)$ will provide an estimate of the reference clock $C_r(t)$ at any time instant $t$.

We explored EPC Gen2 standard and LLRP protocol and found out that LLRP assigns a \emph{FirstSeenTimestamp} in UTC, which is defined as \emph{``The Reader SHALL set it to the time of the first observation amongst the tag reports that get accumulated in the TagReportData''} \cite[p. 87]{llrp}. From this definition, we assumed that this timestamp is assigned by the reader when it receives the EPC during the handshake operation with the corresponding tag, as shown in Fig. \ref{fig:sender_receiver} as the timestamp assigned at time $t_1$. Therefore, \emph{FirstSeenTimestamp} can be considered as $C_r(t_1)$. In order to obtain the corresponding local time $C_w(t_1)$, one strategy is to force reader to send a special ``synchronization'' command after the handshake so that the tag timestamps the command reception event using its local clock, as shown in Fig.\,\ref{fig:sender_receiver} as time $t_2$. The \emph{transmission delay} in this case is $\Delta t= t_2-t_1$ and it is desirable to keep $\Delta t$ as small as possible so that $C_w(t_1)\approxeq C_w(t_2)$. 

\begin{figure}
	\centering
	\includegraphics[scale=0.34]{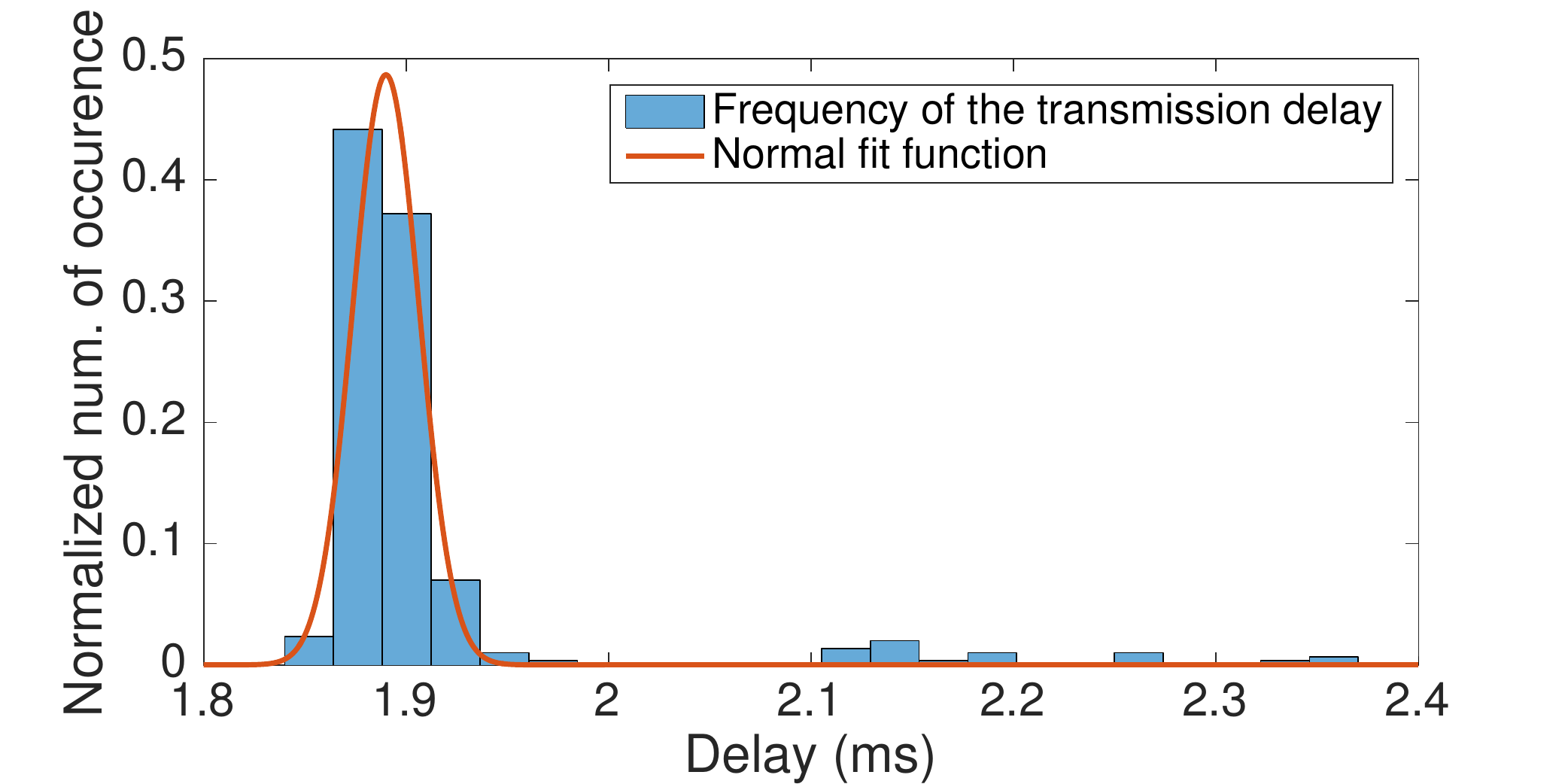}
	\caption{\label{fig:transmission_delay} The normalized number of occurence of the tranmssion delays measured by sniffing the communication between REFID reader and the WISP tag. We calculated the mean transmission delay and its standard deviation as 1.89\,ms and 0.0164\,ms with a 99\% confidence interval of [1.8874,1.8925] and [0.0148,0.0184], respectively.}
	
\end{figure}

\textbf{Transmission delay measurements:} In order to observe and characterize the transmission delay $\Delta t$, we sniffed the communication between the RFID reader and the WISP tag delay during communication scenario in Fig. \ref{fig:sender_receiver} by obtaining 300 samples. Fig. \ref{fig:transmission_delay} presents a summary of our measurements. We observed that, the transmission delay is distributed with a mean of 1.89\,ms and standard deviation of 0.0164\,ms. Even though we observed some outlier values, we think that these are due to the EPC Gen2 implementation of the Impinj reader. After observing the variations of the transmission delay, the next step is to evaluate the limits of sender-receiver synchronization mechanism. To this end, we collected $(C_w(t_2),C_r(t_1))$ pairs for offline processing by controlling the RFID reader to send a \emph{Read} command to the tag and by programming the WISP tag so that it backscatters $C_w(t_2)\approxeq C_w(t_1)$ upon receiving this command. The received pairs $(C_w(t_1),C_r(t_1))$ are logged by the host computer.

\begin{figure}
	\centering
	\includegraphics[scale=0.34]{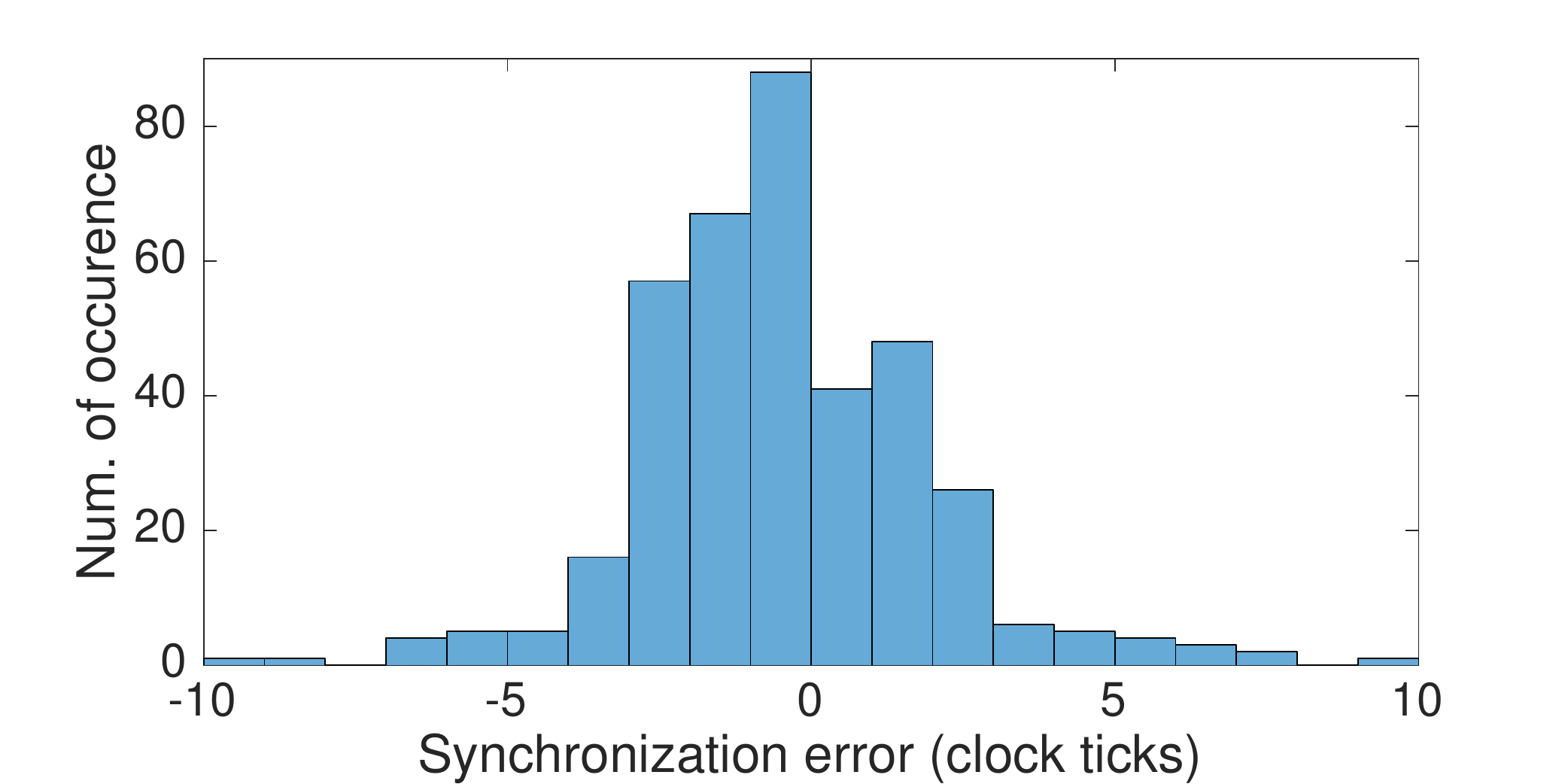}
	\caption{\label{fig:least-squares}Synchronization error by employing least-squares regression on the collected timestamps. We observed a maximum synchronization error of 10 clock ticks between the RFID reader and the WISP tag, leading to 0.32\,ms synchronization accuracy.}
\end{figure}

\begin{figure}
	\centering
	\includegraphics[scale=0.75]{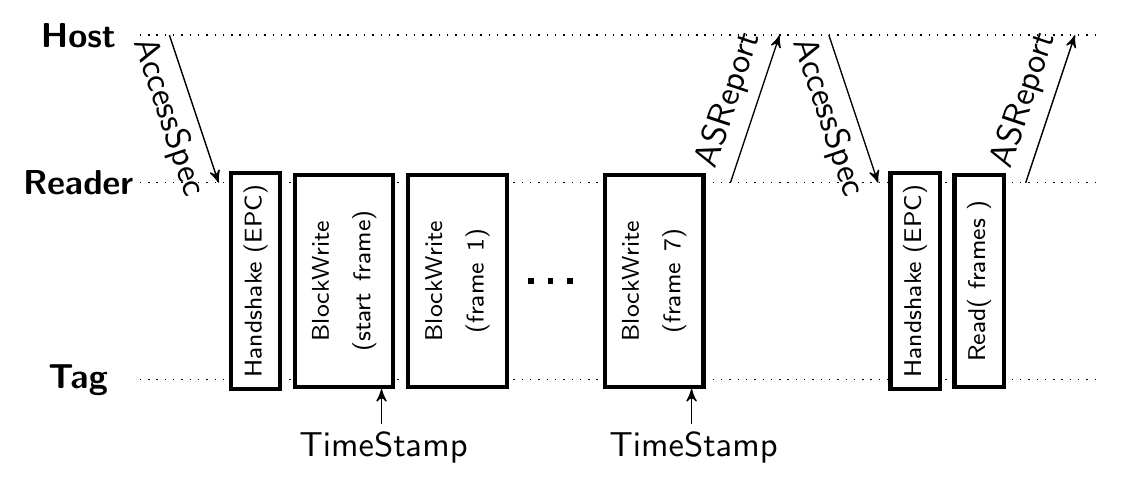}
	\caption{\label{fig:event_based}Event-Based synchronization steps: The WISP tag timestamps successive \emph{BlockWrite} events and adjusts its software clock.}
\end{figure}

\textbf{Software clock computation:} By assuming a linear relationship between $C_r$ and $C_w$, we modeled the software clock of the WISP tag as $$ S_w(C_w(t)) = \alpha + \beta C_w(t) ,$$ where $S_w$ represents an estimator of reader's clock $C_r$, $\alpha$ is the offset and $\beta$ is the relative speed with respect to local clock $C_w$. To establish such a relationship, we performed least-squares regression in MATLAB, as in \cite{Maroti2004}. The idea is that since the WISP tag has limited memory, computation capability and energy, at each step the most recent $N$ pairs are used to calculate the estimated regression line. Formally, let $[C_w(t_k),C_r(t_k)]$ denote the $k$th pair in the log file where $t_k$ denotes the real-time at which \emph{FirstSeenTimestamp} has been assigned during the collection of this pair. At each $k$th step, the pairs $\{[C_w(t_k),C_r(t_k)],...,[(C_w(t_{k+N-1}),C_r(t_{k+N-1})]\}$ are used to calculate the software clock $S_w$. We calculated the synchronization error as $\gamma(t_{k+N})=S_w(C_w(t_{k+N}))-C_r(t_{k+N})$ that represents the difference between the predicted reference time and the received reference time. In our implementation, we used $N=8$ as in \cite{Maroti2004} and Fig. \ref{fig:least-squares} presents the synchronization error at each step. We observed a maximum synchronization error of 0.32\,ms in this one-hop network, which is more than 1 order of magnitude than the synchronization performance of the de facto WSN solution \cite{Maroti2004}, which was reported as approximately 10\,$\mu$s.

\textbf{Limitations of the approach:} In addition to the challenges listed in 
Section \ref{sec:CRFIDs}, we observed two crucial limitations for WISP 
platform, that prevents to build up a sender-receiver synchronization building 
block:

\begin{itemize}
	\item \textbf{Lack of broadcast primitive:} Since RFID reader assigns \emph{FirstSeenTimestamp} for each tag, the synchronization steps in Fig. \ref{fig:sender_receiver} should be repeated for each WISP tag to obtain synchronization in the communication domain of the reader. Unfortunately, in current EPC Gen2 and LLRP standard, we were unable to find any other mechanism to send a global timestamp that is received by all WISP tags inside the communication range of the RFID reader and used to establish synchronization in one step.
	
	\item \textbf{Host computer computation:} We are unaware of any command that will allow to send \emph{FistSeenTimestamp} to the tag. Hence, even though we were able to collect $(C_w(t),C_r(t))$ pairs for offline processing, it is not possible for the WISP tag to collect $C_r(t)$ and synchronize itself to the reader. Therefore, with this limitation, only a host computer can collect and log the timestamps, calculate the relationship between the clock of the WISP tag and the clock of the reader and send the data that represents this relationship to the WISP tag for synchronization. 
\end{itemize} 

\subsection{Approach 2: Event-Based Synchronization}\label{subsec:event_based}

\begin{figure}
	\centering
	\includegraphics[scale=0.34]{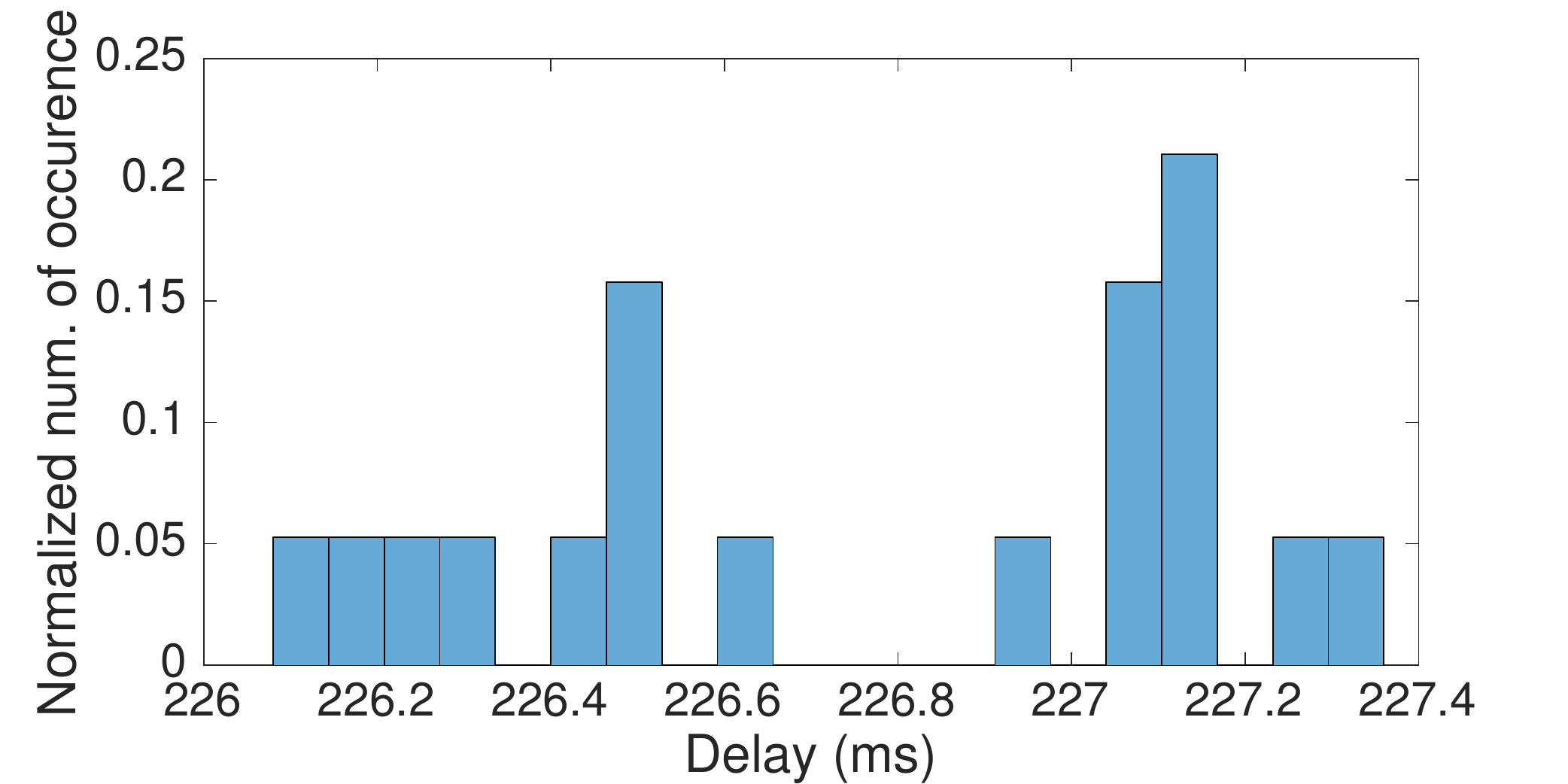}
	\caption{\label{fig:block_write_delay}The delay between the first and the last \emph{BlockWrite} event, i.e. \emph{event period}, by considering 20 samples during the communication scenario in Fig. \ref{fig:event_based}. We measured its mean and standard deviation as 226.7667\,ms and 0.4097\,ms with a 99\% confidence interval of [226.4961,227.0372] and [0.2852,0.6945], respectively.}
\end{figure}

In event-based synchronization mechanisms, a common event which is observable by all receiver devices simultaneously is generated by the reference device. Upon receiving events generated at regular intervals, receiver devices can predict occurence time of the future events. In order to synchronize the WISP tag with such a mechanism, the RFID reader does not send explicit timestamp values as in sender-receiver based synchronization but instead generates events at regular intervals. Upon observing these events, the receiver WISP tag adjusts the rate of its software clock so that it predicts the occurence of the next event precisely. 

\textbf{Observation:} We explored the EPC Gen2 standard to see how to generate events at regular intervals and realized that the \emph{BlockWrite} operation allows us this feature. Fig. \ref{fig:event_based} presents the steps of event-based synchronization. During the command phase, EPC Gen2 allows to perform maximum eight successive \emph{BlockWrite} operations. It is desirable to use the first and the last \emph{BlockWrite} events for synchronization since it is better to compensate frequency differences observed in longer time intervals to adjust the software clock. The real-time length between the first and the last \emph{BlockWrite} operation is the \emph{event period} and its variation is the main error source. Therefore, it is important to explore its charecteristics. 

\textbf{Measurements:} To this end, we sniffed the communication between the RFID reader and a WISP tag presented in Fig. \ref{fig:event_based}. We took 20 sample measurements about the \emph{event period}, which is summarized in Fig. \ref{fig:block_write_delay}. According to our measurements, we observed that the \emph{event period} is distributed with a mean of 226.76\,ms and standard deviation of 0.41\,ms; respectively. 

\textbf{Proposed Solution:} For establishment of the software clock, 
lightweight solutions in terms of computation and memory overhead are crucial 
due to two important requirements: (i) since voltage level is time-varying, 
computations should demand little amount of energy, i.e. marginal number of 
steps, to consume less power (ii) since power is intermittent, the number 
variables pertaining to the software clock should be marginal so that saving 
the state of the synchronization to non-volatile memory will demand less time 
and energy. Considering these facts, we designed a lightweight approach 
inspired from the PI-controller based solution introduced in~\cite{pi2015}, 
which is justified to be an efficient practical solution in WSNs. In order to 
implement our approach, we model the software clock of the WISP tag in the 
real-time interval $[t_0,t] $ as $$ S_w(t) = \alpha(C_w(t)-C_w(t_0) )$$ where $ 
f_{\text{nom}}/f_{\max} \leq \alpha \leq f_{\text{nom}}/f_{\min}$ denotes the 
\emph{rate multiplier } whose value is adjusted to increase or decrease the 
speed of the software clock with respect to the reference clock. Based on this 
model, we introduce Algorithm \ref{alg:sync} that synchronizes the speed of the 
software clock to the reference clock speed using an \emph{integral control} 
strategy. The steps of this algorithm can be explained as follows. Initally, 
the WISP tag 
lets its software clock run at the same speed of its local clock by assigning 
its rate multiplier to 1 (Line 1). Upon receiving the first \emph{BlockWrite} 
event (Line 4), WISP tag stores its local clock value at the variable $t_f$ 
(Line 5). After receiving the last \emph{BlockWrite} event (Line 7), first it 
calculates the difference between the amount of software clock progress and the 
mean event period $\mu_{e}$ (Line 8). Then, it applies the correction on its 
\emph{rate multiplier} $\alpha$ by multiplying the error with the integral gain 
$\beta$ and subtracting it from $\alpha$ (Line 9). After receiving successive 
events, $\alpha$ will converge to its desired value eventually, as can be 
proven by applying similar analytical steps in \cite{pi2015}.

\begin{algorithm}[t]
	\caption{\label{alg:sync}Integral controller-based synchronization algorithm.}
	\small	
	\quad	  \textbf{Definitions:}
	
	\qquad $t_f$: local time of the first \emph{BlockWrite}
	
	\qquad $\alpha$: rate multiplier
	
	\qquad $\mu_{e}$: mean \emph{event period}
	
	\qquad $\beta$: constant integral gain \\
	1: \textbf{Initialization} \\
	2: \quad $\alpha = 1$ \hfill // initialize rate multiplier $\alpha$ \\
	3: \\
	4: $\square$ \textbf{Upon receiving the first \emph{BlockWrite}} \\
	5: \quad $ t_f = C_w(t)$ \hfill // store the local time in $t_f$ \\
	6: \\
	7: $\square$ \textbf{Upon receiving the last \emph{BlockWrite}} \\
	8: \quad $\gamma(t) = \alpha(C_w(t)-t_f)-\mu_{e}$ \hfill // calculate estimation error $\gamma$ \\
	9: \quad $\alpha = \alpha -\beta\gamma(t)$ \hfill // apply integral control to update $\alpha$
\end{algorithm}

\textbf{Convergence conditions:} The selection of the integral gain has significant impact on the performance of the algorithm. In \cite{pi2015}, it has been proven that $0<\beta<2/Bf$ should be satisfied to achieve synchronization in theory where $B$ denotes the event period in seconds and $f$ denotes the clock frequency in Hz. However, smaller values of $\beta$ lead to smaller synchronization error but longer convergence time.

\textbf{Advantages:} From the steps of Algorithm \ref{alg:sync}, it can be observed that this approach employes only very simple arithmetic operations at each step, e.g. Lines 7-9 are composed of only three subtraction and two multiplication operations. Moreover, the \emph{state of synchronization}, i.e. the parameters that are required to be saved in the non-volatile memory, is represented by only the variable $\alpha$. As a consequence, this matches the requirements of the WISP platform. 

\subsubsection{Evaluation of Algorithm \ref{alg:sync}}

In order to evaluate the aforementioned synchronization approach, we considered the value of $\gamma$ in Algorithm \ref{alg:sync}, which represents  the estimation error, as an evaluation metric. We collected the local clock readings of the WISP tag at the first and last \emph{BlockWrite} events and used MATLAB to process the collected data in order to simulate the real behavior of the WISP tag. This gave us flexibility to try different approaches without reprogramming the WISP tag.

\begin{figure}
	\centering
	\includegraphics[scale=0.34]{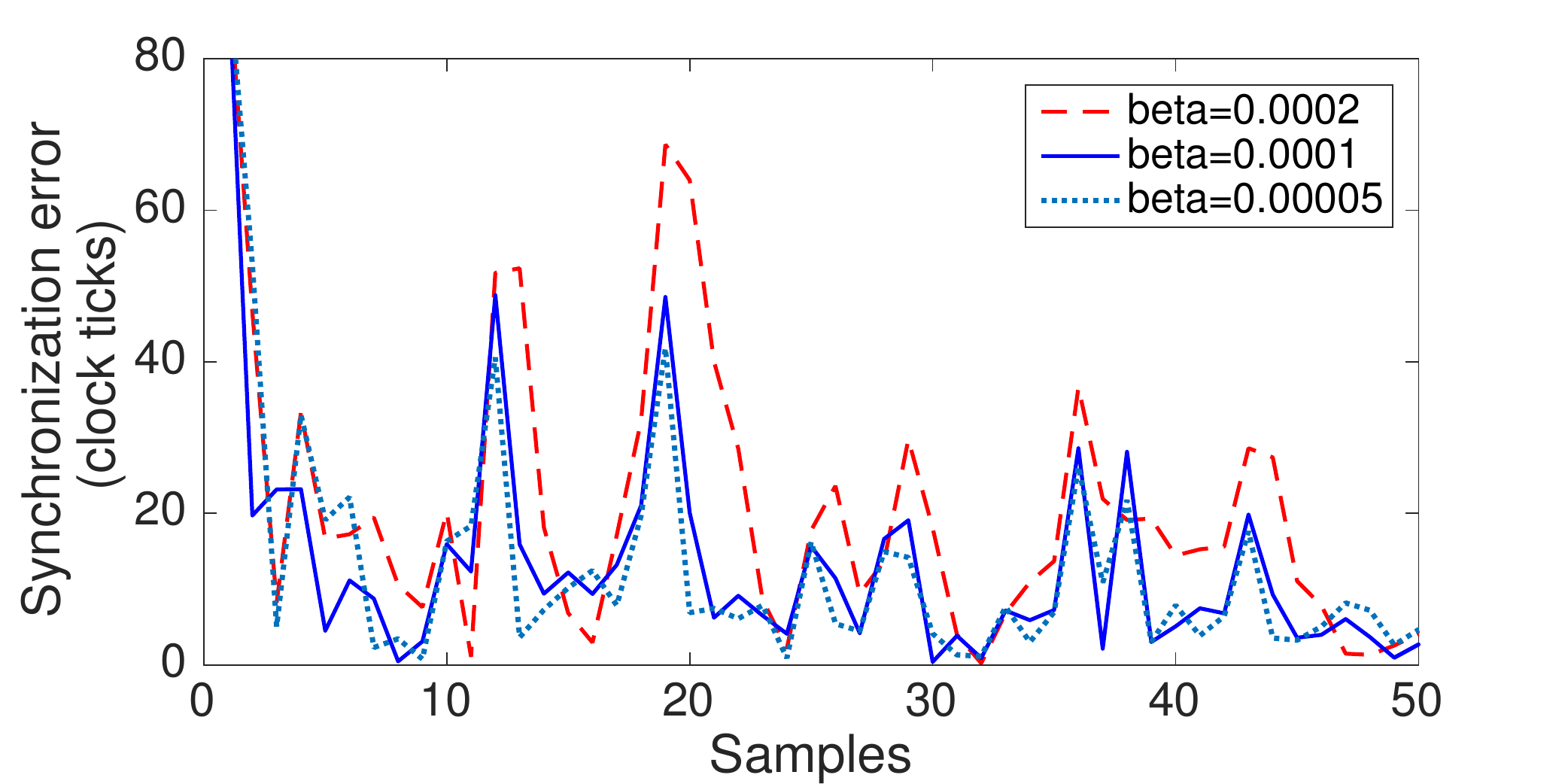}
	\caption{\label{fig:beta} The synchronization error tends to get smaller 
	with smaller integral 
	gains, but after some point deacreasing it has no significant effect due to 
	the precision of 32\,kHz clock.}
\end{figure}

\textbf{Selection of integral gain:} Since we measured the mean event period as 
226.76\,ms on average by experimental evaluation, we set $\mu_{e}=7086$ clock 
ticks since the local clock of WISP tag is operating at 32\,kHz and each clock 
tick occurs at 32 microseconds. First, we explored how $\beta$ effects the 
performance of the algorithm. By substituting $B=0.22676$ sec and $f=32$\,kHz, 
the convergence condition becomes $0<\beta<0.000276$ in our case. In 
consistency with this bound, we observed that synchronization cannot be 
established when $\beta$ is outside this boundary. We present the 
synchronization error with different $\beta$ values in Fig. \ref{fig:beta}. Due 
to the low-precision 32\,kHz clock, the synchronization error tends to get 
smaller with smaller $\beta$ values but after some point, deacreasing $\beta$ 
has no significant effect. Therefore, we chose $\beta=0.0001$ during next 
evaluation steps.

\begin{figure}
	\centering
	\includegraphics[scale=0.34]{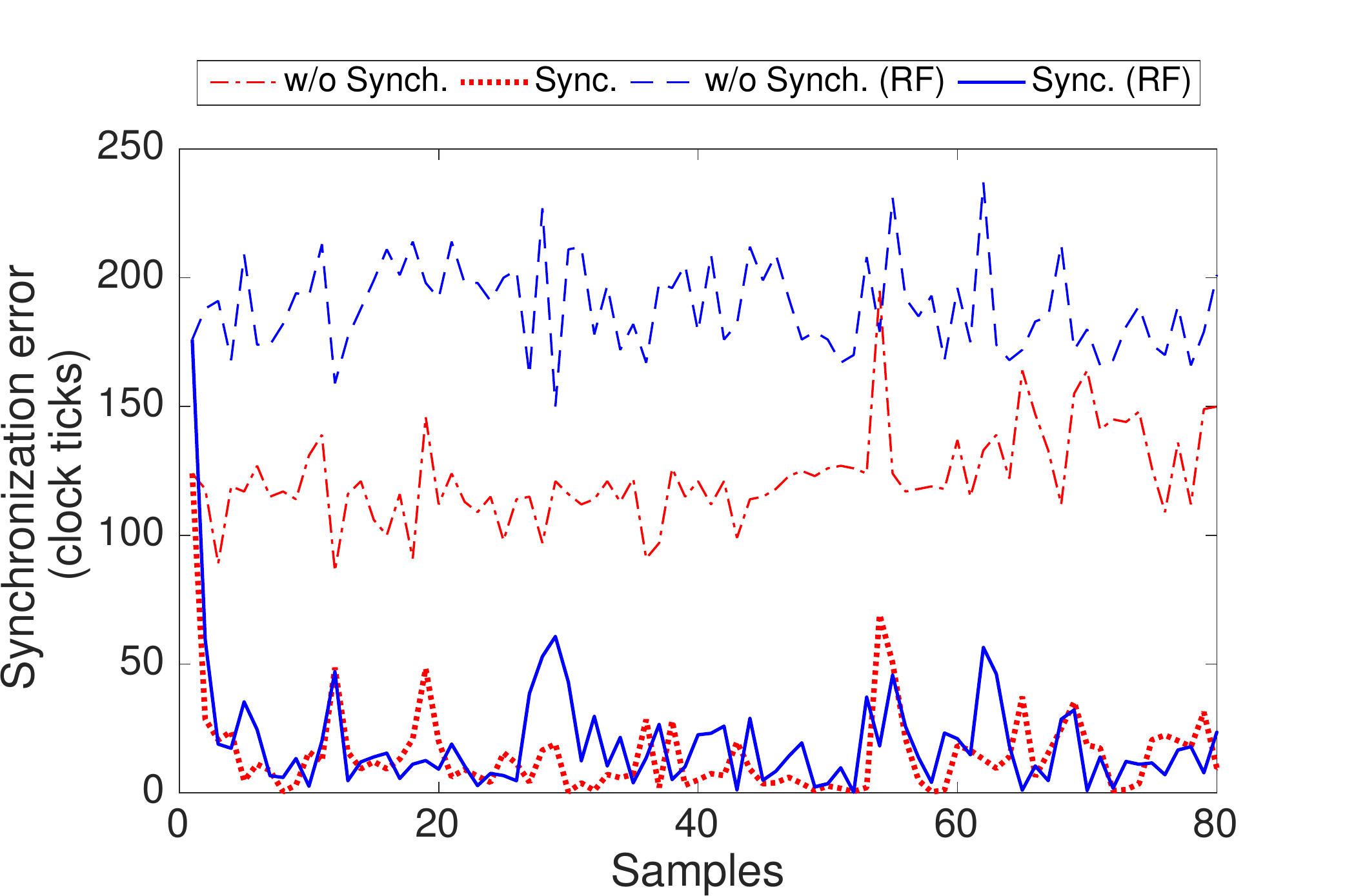}
	\caption{\label{fig:sync_error}Event-based synchronization performance when WISP tag is powered through a constant voltage source (red lines) and through RF power harvesting (blue lines). Mean synchronization errors  without and with Algorithm\,\ref{alg:sync} were almost 127 and 16 clock ticks under stable voltage and they were 175 and 22 clock ticks with RF power harvesting; respectively.}
\end{figure}

\textbf{Synchronization Performance:} Fig. \ref{fig:sync_error} presents synchronization error measurements when the WISP tag is powered by using constant voltage source and powered by only RFID reader through RF waves. Measurements under constant and stable voltage allowed us to observe synchronization under stable clock frequency. We obtained a significant synchronization performance with Algorithm \ref{alg:sync}, almost a factor of 2 less synchronization error as compared to the case where we did not perform any synchronization. It should be noted that even in this case, we observed quite fluctuating synchronization error due to the varying transmission delays, that led a peak error appear between the samples 50 and 60. Apart from experiments with constant voltage input, measurements under highly varying RF power led us to observe the behavior of synchronization under highly varying clock frequencies. In this case, there is a considerable amount of increase on the error, i.e. approximately twice as more, as compared to the stable voltage case. However, we also obtained considerable improvements with our efficient approach with more unstable clock frequency, almost a factor of four better synchronization accuracy.  

\textbf{Limitations of Algorithm \ref{alg:sync}:} In conclusion, experimental 
results indicate that a maximum synchronization error of approximately 1.5 
milliseconds can be ensured between the RFID reader and a WISP tag most of the 
time by employing event-based synchronization mechanism. However, even though 
event-based approach is relatively lightweight and simple as compared to the 
sender-receiver synchronization, the WISP tag is unable to obtain an explicit 
reference clock value. The only variable it tunes is the $\alpha$ that 
represents the relative frequency of the software clock with respect to the 
clock of the reader. Therefore, we require additional steps that will allow 
tags to obtain the actual reference clock value. Moreover, 
Algorithm~\ref{alg:sync} requires knowledge about $\mu_e$, which can be RFID 
reader dependent and challenging to measure it. Last, the steps in Fig. 
\ref{fig:event_based} presents the communication scenario between one WISP and 
the reader. To allow other WISP tags sniff this communication and synchronize 
themselves, broadcast primitive is crucial.

\section{Conclusions and Future Research Directions}

In this paper, we explored a synchronization scenario between an RFID reader and a WISP tag. We studied sender-receiver and event-based synchronization mechanisms in this setting and provided initial designs that will guide future explicit synchronization mechanisms among individual WISPs that reside inside the communication range of a common RFID reader. We provided implementation and evaluation of these designs in our testbed and identified their limitations and drawbacks. Our main finding is to show that with lightweight mechanisms, as of now, a maximum synchronization error of approximately 1.5 milliseconds can be ensured. 

We provide the following issues for future studies in this domain:
\begin{itemize}
	\item \textbf{Network-wide synchronization:} Currently, we provided synchronization between an RFID reader and a single WISP tag. However, synchronization among all WISP tags inside a single communication domain and also synchronization of the whole CRFID network composed of several RFID readers and WISP tags is still an issue, due to the single-hop nature of backscatter communication. Since WISP tags can only communicate with the reader, not with their neighboring tags, it is also interesting to study \emph{wisp-wisp} synchronization and explore its feasibility. 
	
	\item \textbf{Design of a broadcast primitive:} As we discussed in Section \ref{subsec:sender_receiver}, lack of a broadcast primitive for the WISP platform limited our designs and implementations. We think that broadcast primitive is crucial not only for synchronization but also for other protocols and applications in the CRFID domain.
	
	\item \textbf{Power loss and recovery:} It is crucial to save the synchronization status to non-volatile memory, just before power loss. However, it is not the only system state to be saved and writing to non-volatile memory also consumes energy. Hence, when and how to save the synchronization status is worth to explore. 
	
	\item \textbf{Voltage-frequency relations:} It might be interesting to explore the voltage-clock frequency relationship since WISP tags are subject to varying voltage source when they are powered with only RF energy harvesting from the RFID reader. We anticipate that voltage level should be incorporated to the establishment of the software clock so that voltage dependent instability of clock frequency can be compensated. However, this is a ``chicken or eg'' problem since reading voltage level also consumes energy. 
	
	\item \textbf{Synchronization in other IPD platforms:} Synchronization in other IPD platforms, such as embedded systems that use ambient backscatter communication \cite{ambient_backscatter}, is worth to explore as well. 
\end{itemize}

 %An important requirement, especially for the testbed to evaluate the synchronization protocols, is the collection of instantaneous clock values from individual WISPs that allows to obtain the maximum instantaneous synchronization error among WISP tags. This requires generation of a common event so that WISP tags save their clock values at the event occurence. Such common event generation also requires a kind of broadcast capability. 

\bibliographystyle{abbrvnat}
\bibliography{references}

\end{document}